# Superconducting Non-Contact Device for Precision Positioning in Cryogenic Environments

J.L. Perez-Diaz, *Member, ASME*, I. Valiente-Blanco, E. Diez-Jimenez and J. Sanchez-Garcia-Casarrubios.

*Abstract*—In this paper, a non-contact linear positioner based on superconducting magnetic levitation for high-precision positioning has been tested under cryogenic conditions (~20 K and ~$10^{-6}$ Pa). The prototype is able to achieve submicrometric positioning resolution of 230 ± 30 nm RMS along a stroke of ± 9 mm length with a current resolution of 15 µA, and a peak current requirement lower than ± 500 mA. In addition, it was demonstrated that an open-loop control strategy could be used for positioning the moving part with accuracy of the order of 1 µm. On the other hand, deviations of the slider position were found to be ± 650 µrad for the pitch, lower than 100 µrad for the yaw, ± 2000 µrad for the roll and ± 4 µm for the lateral run, all of them related to a full stroke motion. These results reveal a good performance of the device and demonstrate the potential of a new tool for applications where high-precision positioning is required within a long range in cryogenic environments like far-infrared interferometry.

I. INTRODUCTION

In recent years, precision positioning has become an important development target for meeting the requirements of the semiconductor, optical communication, chemistry, assembly, manufacturing process, bio-medical and precision industries. Given this context, the development of a long-range nanometric positioning stage has become a hot research objective [1].

In particular, far infrared interferometers are currently intended to be used on satellites for detecting earth-type planets and for researching the origins of galaxies, planetary systems and stars. A mirror that can be positioned with extreme precision along a large stroke must also be provided [2] as a part of these types of instruments. It is noteworthy that the lower the temperature that the infrared sensor operates at, the higher sensitivity the device will have [3]. This makes cryogenic conditions a mandatory requirement for these kinds of devices. In addition to the cryogenic environment, very low energy consumption is also desirable.

Usually, piezoelectric actuators are chosen to design precise positioning systems in order to achieve the submicrometric or nanometric positioning. Articles and patents related to piezoelectric actuators are numerous [4–7].

Nevertheless, piezoelectric actuators have several limitations, for example, they are sensitive to environmental changes such as temperature [8]. Nonlinearities such as hysteresis and creep are also present in piezoelectric materials [9–11]. Furthermore, the voltage required to operate piezoelectric actuators can be as high as several-hundred volts, which can be a problem in situations where it is necessary to have low levels of electrical input. However, the main disadvantage of piezoelectric actuators remains the very limited motion range, usually not longer than a few hundred micrometers [12], [13]. Despite these limitations, piezoelectric-based actuators are widely used nowadays for applications where highly accurate positioning is required within a short range.

Hybrid or dual-stage positioning systems based on piezoelectric actuators can sometimes reach long strokes with impressive accuracy. Such hybrid devices combine one actuator for a coarse step motion, and piezoelectric actuators to achieve a high level of precision, in the order or a nanometers [14], [15]. Currently, the friction type piezoelectric stage can be integrated with a linear motor system to achieve a resolution in the nanometer range and a motion stroke of even more than 10 mm length. This is also available for use in industrial applications [16].

A very high precision positioning system for a fiber-based confocal microscope for cryogenic spectroscopy using piezoelectric materials was proposed by Hoegel in 2008. A motion range of 7.5 mm, a resolution of 5 nm and suitability for temperatures of 4.2 K were reported [17].

This work has been partially funded by D. G. Economía, Estadística e Innovación Tecnológica, Consejería de Economía y Hacienda, Comunidad de Madrid, ref. 12/09.

J.L.Perez-Diaz, Departmento de Ingeniería Mecánica, Universidad Carlos III de Madrid, Leganés E28911 SPAIN (phone/fax: +34916249912; e-mail: jlperez@ing.uc3m.es).
I.Valiente-Blanco, Departmento de Ingeniería Mecánica, Universidad Carlos III de Madrid, Leganés E28911 SPAIN; e-mail: ivalient@ing.uc3m.es).
E.Diez-Jimenez, Departmento de Ingeniería Mecánica, Universidad Carlos III de Madrid, Leganés ( e-mail: ediez@ing.uc3m.es).
J.Sanchez-Garcia-Casarrubios Departmento de Ingeniería Mecánica, Universidad Carlos III de Madrid, Leganés E28911 SPAIN (e-mail: jusanche@ing.uc3m.es).

Other researchers have focused their efforts on magnetic levitation (maglev) positioners. In 1996, Holmes and Trumper published the design of a magnetic fluid bearing system to achieve a stroke of 100 µm and a resolution of 0.1 nm [18]. Kim and Trumper developed and demonstrated a high-precision planar maglev stage with 50 mm planar motion capability, which had a 10 nm resolution and 100 Hz control bandwidth [19]. These new kinds of devices eliminated the strain, backlash, and hysteresis that limited the precision of position control. In 2000, a maglev scanning stage with a 0.6 nm three-sigma horizontal position noise was built and demonstrated [20]. Two years later, in 2002, Shan et al. published a new proposal for a multiple degrees of freedom magnetic suspension stage [21].

More recently, active maglev stages with greater travel ranges and the same resolution can be found. For example, there is a 6 degrees of freedom active maglev system which has been improved for 5 years by Won-Jong Kim, Shobhit Verma et al. between 2004 and 2007, achieving a travel range of 5 mm in each direction and 3 nm of resolution[22], [23]. Also a 2 mm x 2mm x 2mm 6 DoF system was developed by Zhang and Menq in 2007 with c.1 nm RMS resolution [24]. Van den Dool et al. (2009) mounted an interferometer mirror on an active linear magnetic bearing. He reported a stroke over 20 mm with 0.5 nm accuracy at 25 K [25].

In comparison with the piezoelectric positioners, the active magnetic levitation systems can offer the same accuracy for longer motion ranges. They can also offer movement in more than one degree of freedom [26], [27]. By being contactless devices, long lifetime and fatigue problems are minimized. Nevertheless, the main disadvantage of these active magnetic levitation devices is that a great effort is required to control them as they are naturally unstable. This requires a complex control method and consequently, complex electronics. Greater electrical consumptions in a cryogenic environment cause extra heat and therefore additional cooling power is needed.

In this context, neither devices based on piezoelectric materials (due mainly to their short motion travels) nor active magnetic levitation devices (due to the complexity of control and great energy consumption) seem to be suitable to satisfy the requirements of long range nanopositioning at low temperatures, as required for far-infrared interferometers on satellites.

Superconductors provide inherent stability that greatly simplifies the control strategy and, as they require very low temperatures to operate, they are very suitable for cryogenic environments. The idea of using superconductor levitation for contactless devices has also been previously explored [28]. However, the choice of shape, configuration and size of the superconductors is not a small issue and requires analytical and experimental studies [29–32]. In 1989, A. Masatake et al. claimed the design of a superconducting actuator that consists of a planar motion device that floats over a matrix of electromagnets. This matrix can be set up to move and place the levitation device in certain positions. No tests or results about position accuracy were presented [33].

Some kinds of conveyors have shown reasonable results. In 1997, Iizuka and Fujita published a conveyor with 2.8 mm stroke and an accuracy of 40 µm [34]. In 2009, they published a similar device with an accuracy within millimeters [35]. In 2007, Lin et al. modified the height of levitation of a permanent magnet over a field cooled superconductor bulk by changing the operating current in a properly located Helmholtz coil. Maximum precision obtained by the authors was in the range of µm and the total stroke reached in this experiment was around 140 µm [36].

Some other patent claims have been made in recent decades. In 1993, Goodon et al presented a floating adjustable device based on the concept of a magnet floating over a matrix of superconductors [37]. The control system is made by applying a voltage difference directly over the superconductors to move the permanent magnet. A combination of north and south poles generated into the superconductor part is what moves the floating permanent magnet. The precision with which the magnet is positioned was not discussed. Another interesting patent was applied by G.R. Brotz in 1991 [38], who claimed for an apparatus involving a moving mirror that used the Meissner superconducting effect in order to stabilize a mirror in a floating, non-accurate position. Once this had been achieved, it was proposed that the mirror would move accurately using different coils placed near to it.

A linear positioning device based on superconducting magnetic levitation is presented in this paper. A large stroke of 18 mm was achieved with submicrometric resolution using a non-contact magnetic actuation system that proved very easy to control. This new approach to high-precision positioning in cryogenic environments based on superconducting magnetic levitation has led to very promising results, obtaining similar performance as other technologies, but with the benefit of a much simpler open loop control strategy.

## II. DEVICE DESCRIPTION

The device studied in this paper is a non-contact linear slider based on superconducting magnetic levitation. The invention [39] is mainly composed of a static guideline, made of two 45 mm diameter superconducting polycrystalline $YBa_2Cu_3O_{7-x}$ disks (HTS) (1), and a slider (2) composed of a long $Nd_2Fe_{14}B$ permanent magnet (PM), 160 mm in length, as shown in Fig. 1 (a). In order to move the slider along the path, magnetic forces are exerted on the PM using magnetic fields generated by two coils (3) placed at either end of the stroke as shown, for example, in Fig.1.b).

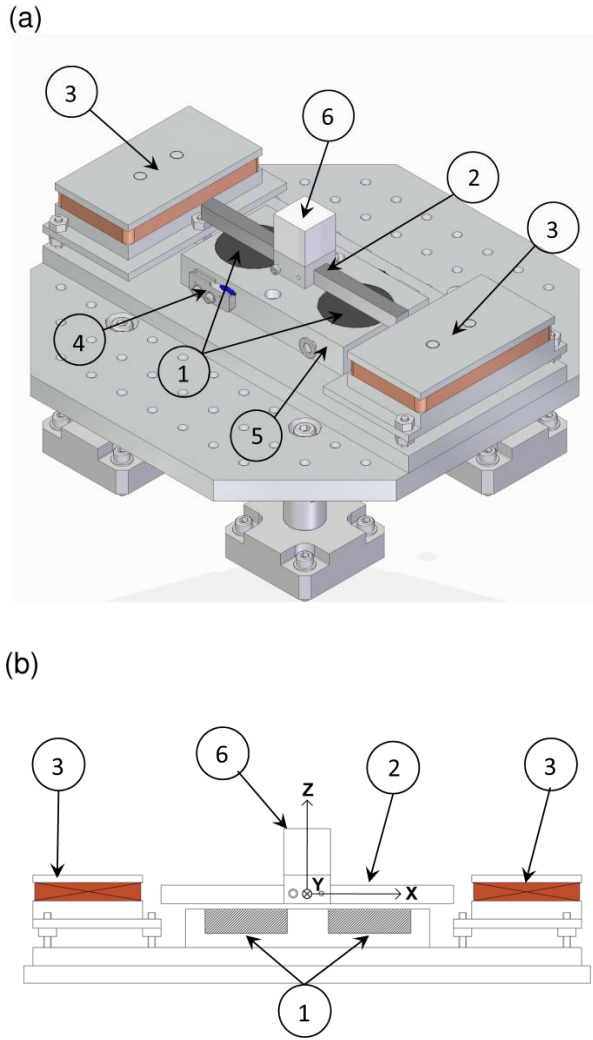

Fig. 1. 1 YBaCuO superconductor disks (HTS); 2 long permanent magnet (PM); 3 aluminum core coils, 4 PT-100 sensors; 5 aluminum vessel for the HTS and 6 aluminum polished optic cube. (a) CAD representation of the device; (b) Lateral representation. Reference system used in this paper.

The reference system used throughout this paper is shown in Fig. 1. (b), which is a lateral cross section of the device shown in Fig. 1. (a).

The magnetization direction of the PM was oriented vertically, i.e. parallel to the Z axis in Fig. 1. (b). The PM was carefully placed with its longitudinal axis coincident with the X axis of the reference system. After that, the HTS were cooled down by immersion in the magnetic field (field-cooled) generated by the PM. Once the superconductors are at a temperature lower than their critical temperature (Tc ≈ 90 K) [40], the magnetic flux trapped in the superconductor makes the PM levitate stably over the superconductor disks without any contact [41].

From magnetostatics [42], drag forces exerted by the PM on the HTS can be calculated by the integral over their volumes as show in eq. 1:

$$F = \int_V (M \cdot \nabla) B \ dV \quad (1)$$

Where $B$ is the magnetic flux density and $M$ and $V$ are the magnetization and the volume of the superconductor respectively.

Due to a high translational symmetry of the magnetic field seen by the HTS [29] for any X positions of the PM, the HTS and the PM form a kinematic pair such that a "sliding path" is established in the X direction. The previously mention "sliding path" can be intuited in Fig. 2.

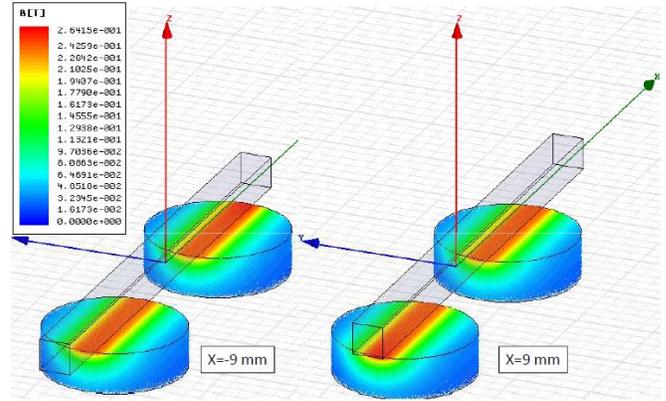

Fig. 2. Magnitude of the magnetic field (Teslas) applied on the superconductors for different X positions of the slider for a height of levitation of 3mm between HTS top face and the PM bottom face.

Thus, the PM can move along the X direction with very little resistance, provided the ends of the PM are far enough from any edge of the superconductors. However, if any end of the PM moves up any edge of the HTS, the symmetry of the magnetic field is broken. Then, due to the variation in the magnetic field applied on the HTS, drag forces will increase [29], [31], [43], trying to restore the slider to its initial equilibrium position at X=0 mm.

Due to symmetry of the mechanism about the XZ plane in Fig.1 (b), the force exerted on the PM by the HTS in the X direction can be calculated as show in eq.2:

$$F_x = -\int_V \left\{ M_x \cdot \frac{\partial B_x}{dx} + M_z \cdot \frac{\partial B_x}{dz} \right\} dV \quad (2)$$

Therefore, the force exerted on the permanent magnet due to its interaction with the HTS is determined by the magnetization of the HTS and the gradient of the X component of the magnetic field. The absolute value of the X component of the magnetic field applied on the superconductors |Bx| at the normal state is shown in Fig. 3.

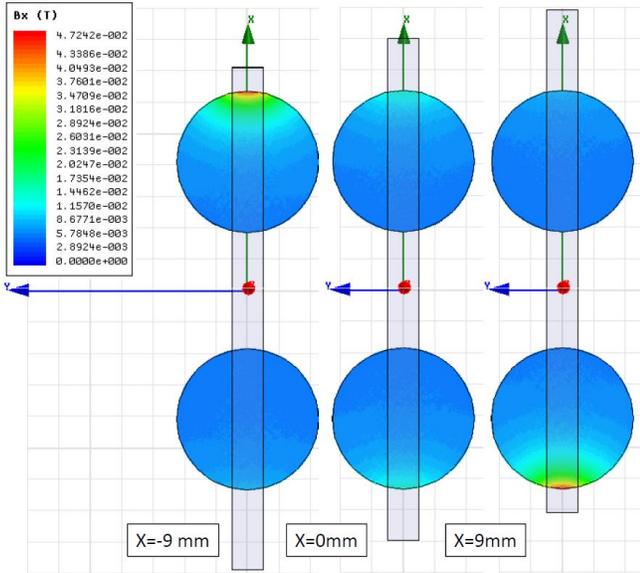

Fig. 3. Absolute value of the magnetic field (Tesla) applied on the superconductors in the X direction for different X positions of the slider for a height of levitation of 3mm between HTS top face and the PM bottom face.

The performance of the device was demonstrated to depend on the geometric parameters of the mechanism such as the distance between HTS or the height of levitation of the PM over the HTS. In addition, the equilibrium position of the slider can be modified by applying magnetic forces generated by a pair of coils placed at either end of the stroke [29], [44]. Size and shape of the coils were designed to reach a full stroke motion with a current lower than 1A. Therefore the X position of the slider can be open-loop controlled by the current circulating through the coils. Ultimately, it is relevant that the device presents much stronger rigidity in any other degree of freedom due to the symmetry configuration of the magnetic field seen by the HTS, which remains almost constant for the motion of the slider in the X direction but strongly changes when the PM is moved in any other degree of freedom.

### III. EXPERIMENTAL SET UP

The device was tested in a high vacuum ($2 \times 10^{-8}$ Torr~$2.7 \times 10^{-6}$ Pa). The whole mass of the prototype was cooled down to a working temperature of around $20 \pm 2$ K- much lower than the superconductor's critical temperature (90 K). It has to be mentioned that the temperature of the HTS was constantly being monitored using two Class-A PT-100 sensors (4) (see Fig. 1.a).). The maximum temperature under operating conditions was always well below the HTS critical temperature. The distance between the centers of the HTS disks was set to 84 mm using an aluminum vessel specifically designed for this task (see Fig. 1.a).). Several indium layers were used in order to ensure good thermal contact between the aluminum base of the HTS, the PT-100 sensors and the HTS.

A launch and lock system, fixed to the cryostat base plate, was used to keep the permanent magnet at the initial cooling symmetric position and released or locked when necessary. The levitation height was established to 3 mm between the HTS upper surfaces and the bottom surface of the PM.

The circulating current in the coils was generated by a 16 bit NI 6230 voltage generation device and amplified before supplying the coils. Therefore, the current circulating through the coils was established by the resultant impedance of the electric circuit. This impedance was mainly set by high precision resistors. In this way, the addition of input resistors improves the current resolution without a significant loss of precision. Hence, a better motion resolution could be achieved. Results of this procedure will be discussed in depth in section IV.A. Finally, the circulating current in the coils was measured with a NI 4070 FlexDMM multimeter of 1µA resolution and a 10 p.p.m. accuracy at full scale (1 A).

In order to measure the position of the slider, an Agilent 10706B High Stability Plane Mirror Interferometer with a resolution of 0.62 nm was placed outside the cryogenic environment. A Newport collimator, model LDS Vector (accuracy to within 3%) was used to measure rotations of the slider. Lastly, a polished aluminum optical cube (6) was attached to the slider, as shown in Fig. 1. It was used to reflect the laser beam required by both the interferometer and the collimator.

### IV. RESULTS AND DISCUSSION

#### A. X positioning performance

The position of the slider versus the current circulating in the coils was measured along the whole stroke of the mechanism ($\pm$ 9 mm). The results of these measurements for the full stroke range are discussed below and summarized in Table I in the conclusions section.

*Full stroke motion*

The circulating current in the coils vs. X position of the slider is shown in Fig. 4. Multiple full cycles ($\pm$ 9 mm) were measured. The current in the coils was modified in increments of around $11.2 \pm 0.4$ mA. The standard deviation for every X position in Fig. 4 was always lower than $\pm 0.25$ µm and the accuracy in the current signal was always lower than 50 µA.

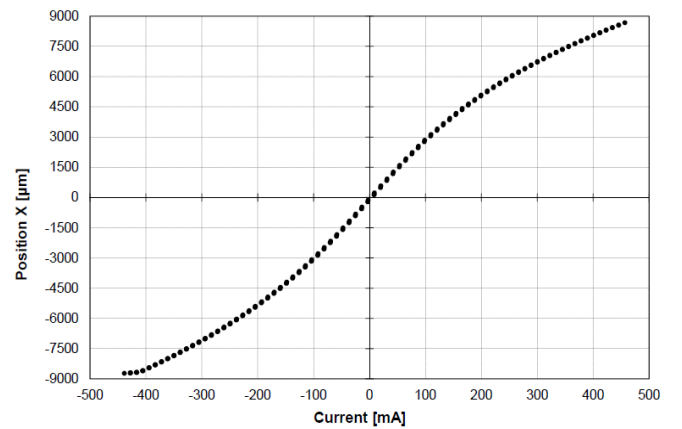

Fig. 4. X position of the slider vs. current in the coils. $\sigma_{XPos} = \pm 0.25$ µm and $\sigma_{Current} = \pm 50$ µA.

Fig. 4 shows that the motion of the slider is not perfectly linear, rather it presents an increasing resistance to motion for higher values of X position. As explained in a previous section, this behavior is related to the breaking of the symmetry of the magnetic field seen by the HTS. It can also be seen that the stroke reached by the device was about ± 9 mm and the current needed to achieve this stroke was around ± 450 mA. Although the sensitivity of the motion was not constant along the entire stroke, it remains equal to 29 ± 2 nm/µA in a range of 6 mm around the central equilibrium position.

*Position stability and accuracy*

X position and Y position of the slider vs. time around the initial equilibrium position (X=0 mm) are shown in Fig. 5. Stability around other X position showed a similar behavior. Consequently, the X position accuracy can be estimated to be of the order of 1µm ($\sigma_{XPos}$= ± 0.25 µm). On the other hand, the accuracy in the Y position can be estimated to be 6 µm ($\sigma_{XPos}$= ± 1.2 µm).

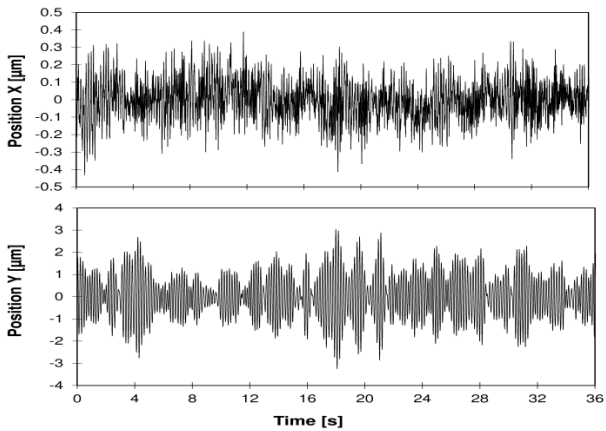

Fig. 5. Stability of the X position (top) and Y position (bottom) of the slider vs. time in the central position (no current in the coils). Acquisition frequency of 1 kHz.

*Motion resolution*

By adding a second input current signal to one of the coils, a finer motion control could be achieved. This second input current was generated as a differential voltage signal, and transformed into current increments (ΔC) by increasing the equivalent impedance of the coils' electrical circuit. Then, the minimum current increment was modified from around 100 µA to 15 µA. The accuracy of the current measured was 1 µA.

*- Coarse Step Resolution*

The X position of the slider vs. current step is represented in Fig. 6. The position resolution under this operating condition was calculated to be 1.5 ± 0.2 µm.

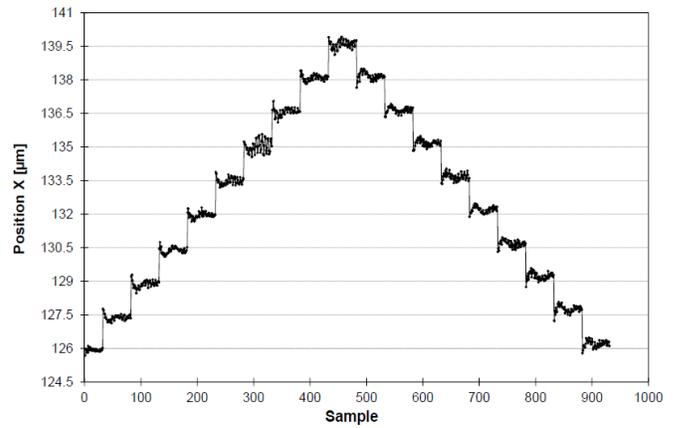

Fig. 6. X Position of the slider vs. sample for ΔC=100±1 µA. Samples measured consecutively with no registered time.

*- Fine Step Resolution*

In Fig. 7, the X position (RMS) of the slider is represented for each current increment (ΔC) in one of the coils. The X position resolution in this case was calculated to be 230 ± 30 nm.

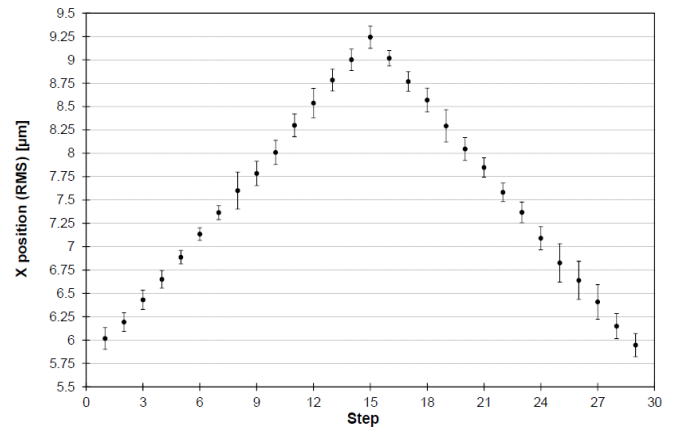

Fig. 7. X position (RMS) of the slider vs. current step for ΔC=± 15±1 µA. Samples measured consecutively with no registered time.

The X position of the slider vs. current in the coils is represented in Fig 8. The sensitivity of the fine step motion was calculated to be 15 ± 2 nm/µA for a ± 3 mm travel range. This is half of the coarse step motion sensitivity because only one coil is used.

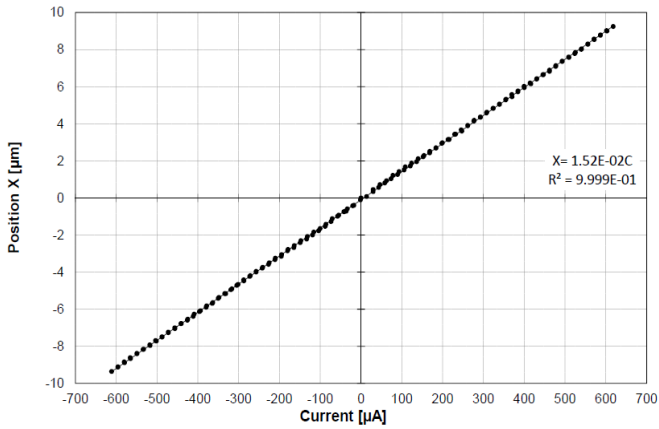

Fig. 8. X Position of the slider vs. current for ΔC=15±1 µA.

*Hysteretic behavior*

The hysteretic behavior is thought to be mainly related to changes in the magnetization of the HTS when they are at the mixed state and the presence of defects in the crystalline structures and chemical impurities in the superconductors that prevent the free mobility of the vortex in the superconductor [45]. In the case of the device introduced in this paper, the hysteresis was calculated to be around 60 µm for a full stroke motion. This hysteretic behavior is shown in Fig. 9, where a zoom has been applied in the surroundings of the origin of coordinates in Fig. 4. It may be noticed that, at this point, the hysteresis was found to be the maximum.

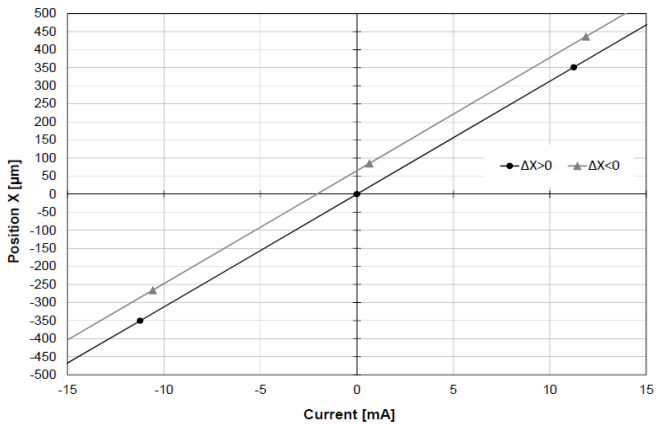

Fig. 9. Hysteretic behavior of the X position of the slider. $\sigma_{XPos} = \pm 0.25$ µm and $\sigma_{Current} = \pm 50$ µA.

B.  *Lateral run out*

The lateral run out (deviation in the Y axis) vs. X position of the slider is shown in Fig 10. The run out was around ± 4µm for a ± 9 mm stroke. It is remarkable that the run out is three orders of magnitude lower than the stroke of the mechanism. The hysteretic behavior in the Y run out is also noteworthy, as can be seen in Fig. 10

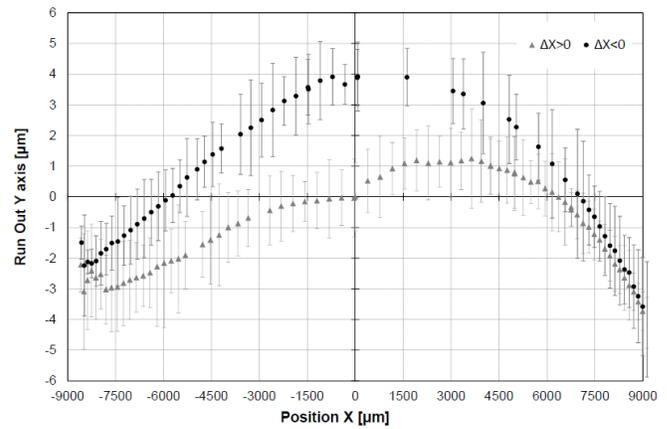

Fig. 10. Lateral run out vs. X position of the slider. $X_{Pos.}$ accuracy= 50 µm.

C.  *Pitch*

Pitch (rotation around Y axis) vs. X position of the slider for the full stroke was measured. The results from these measurements are shown in Fig. 11. The maximum and minimum pitches were obtained close to the end of the stroke and they had a value around ± 650 µrad. Pitch is assumed to be related to the action of gravity and the unbalancing of the PM when it is moved away from the symmetric initial position (X=0 mm). Once again, hysteretic behavior in the pitch can be observed in Fig. 11.

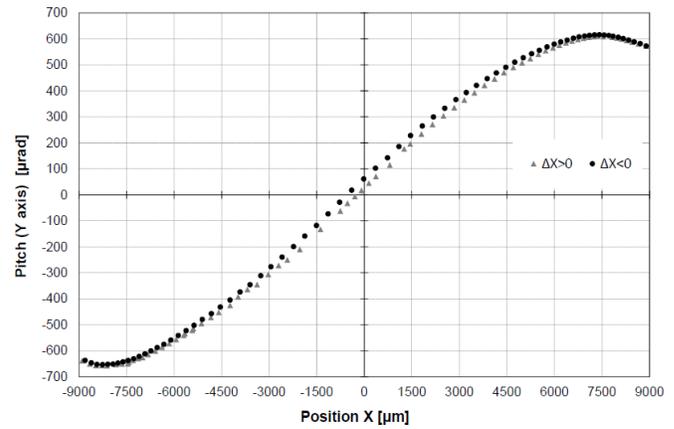

Fig. 11. Pitch (Y rotation) vs. X position of the slider. $X_{Pos.}$ accuracy= 50 µm.

D.  *Yaw*

Yaw (rotation around Z axis) vs. X position of the slider for the full stroke motion is shown in Fig. 12. It can be derived from Fig. 12 that the maximum relative Yaw for a full stroke motion is always lower than 110 µrad. Again, one can observe clear hysteretic behavior in the yaw.

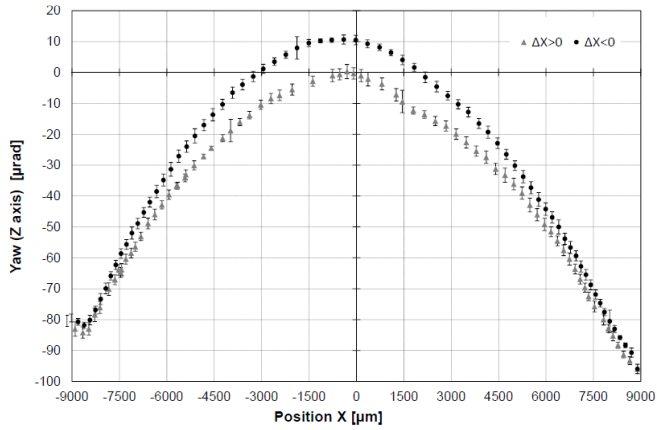

Fig. 12. Yaw (Z rotation) vs. X position of the slider. X $_{Pos.}$ accuracy= 50 µm.

*E. Roll*

Roll (rotation around X axis) vs. X position of the slider for a full stroke motion is shown in Fig. 13. Roll was always between ± 2000 µrad, with a clear hysteretic behavior. Both, roll and yaw were found to be related to initial misalignments between the axis of the coils and the magnetization axis of the PM.

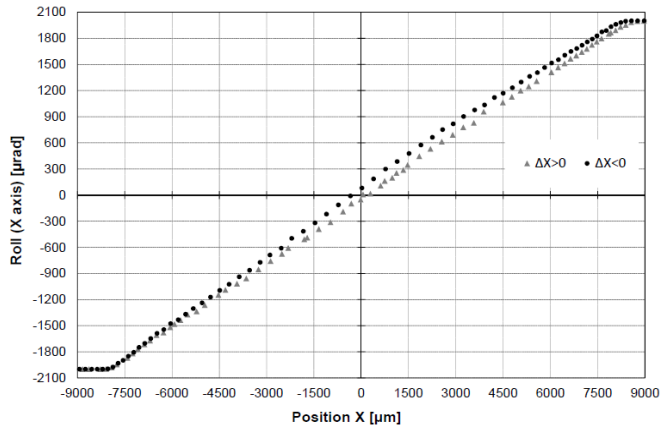

Fig. 13. Roll (X rotation) vs. X position of the slider. X $_{Pos.}$ accuracy= 50 µm.

## V. CONCLUSION

In this paper, it has been demonstrated that a device based on superconducting magnetic levitation achieved submicrometric positioning resolution in a long range motion of ± 9 mm.

The position of a long PM levitating over two HTS at the mixed state can be easily controlled with a very simple open-loop control strategy, by means of regulating the current in two coils placed at either end of the stroke. This discovery not only significantly simplifies the motion control with respect to other magnetic positioning systems, but also reduces the total power consumption and heat dissipation. Both of these issues are particularly important in cryogenic environments and space applications like far-infrared interferometry. The performance of the device presented in this paper is summarized in Table I.

Table I. Summary of the performance of the device

| | |
|---|---|
| **Stroke [mm]** | ± 9 |
| **Sensitivity Coarse / Fine Step motion in the 8 mm linear stage [nm/µA]** | (29 ± 2)/(15 ± 1) respectively |
| **Resolution Coarse [µm]** | 1.5 ± 0.2 |
| **Resolution Fine [nm]** | 230 ± 30 |
| **Position Stability X [µm]** | 1 |
| **Hysteresis of X motion at the central position (X=0) [µm]** | 60 |
| **Lateral Run Out (Y axis) [µm]** | ± 4 |
| **Relative pitch (Y axis) [µrad]** | ± 650 |
| **Relative yaw (Z axis) [µrad]** | < 100 |
| **Relative roll (X axis) [µrad]** | < 4000 |

Besides that, the motion resolution achieved in the linear stage with a 15µA current resolution was 230 ± 30 nm. Nevertheless, it was demonstrated that a better resolution could be achieved by simply improving the current resolution of the input signal. Additionally, the use of a double coil for coarse/fine magnetic field generation could also improve the resolution.

On the other hand, the accuracy of X positioning was of the order of 1 µm, mainly limited by the presence of vibrations. Despite this inconvenience, the device showed great potential to be improved in terms of accuracy as well as resolution. Finally, the measured lateral run out (Y axis) was three orders of magnitude lower than the stroke of the device and the rotation deviations are reasonably low as well, as shown in Table I. In addition, it is expected that rotation deviations could be improved providing a better alignment of the slider in the initial position.

### ACKNOWLEDGMENT

The authors would like to offer their heartfelt thanks to LIDAX for their technical support and cooperation in this project. We would especially like to thank Javier Serrano, Fernando Romera, Heribert Argelaguet-Vilaseca and David González-de-María for their invaluable support. We also like to thank Raghu Menon from Glasgow University and Karen Bernstein for their generous help improving this paper.